\documentclass[english,aps,prl,twocolumn,superscriptaddress]{revtex4-2}
\usepackage[T1]{fontenc}
\usepackage[latin9]{inputenc}
\usepackage{color}
\usepackage{babel}
\usepackage{amsmath}
\usepackage{amssymb}
\usepackage{graphicx}
\usepackage[unicode=true,pdfusetitle,
 bookmarks=true,bookmarksnumbered=false,bookmarksopen=false,
 breaklinks=false,pdfborder={0 0 1},backref=false,colorlinks=false]
 {hyperref}

\makeatletter
\newcommand{\Tr}{\mathrm{Tr}}
\usepackage{braket}
\usepackage[caption=false]{subfig}
\usepackage{algorithm,algpseudocode}
\usepackage{amssymb}
\usepackage{tikz}
\usepackage{adjustbox}

\makeatother

\begin{document}
\global\long\def\k#1{\Ket{#1}}%
\global\long\def\b#1{\Bra{#1}}%
\global\long\def\bk#1{\Braket{#1}}%
\global\long\def\Tr{\mathrm{Tr}}%
\global\long\def\Var{\text{Var}}%
\global\long\def\renyi{\text{R}\'e\text{nyi}}%

\preprint{This line only printed with preprint option}
\title{Superselection-Resolved Entanglement in Lattice Gauge Theories: A
Tensor Network Approach}
\author{Noa Feldman }
\affiliation{\textsuperscript{}Raymond and Beverly Sackler School of Physics and
Astronomy, Tel-Aviv University, Tel Aviv 6997801, Israel}
\author{Johannes Knaute }
\affiliation{Racah Institute of Physics, The Hebrew University of Jerusalem, Jerusalem
91904, Givat Ram, Israel}
\author{Erez Zohar}
\affiliation{Racah Institute of Physics, The Hebrew University of Jerusalem, Jerusalem
91904, Givat Ram, Israel}
\author{Moshe Goldstein }
\affiliation{\textsuperscript{}Raymond and Beverly Sackler School of Physics and
Astronomy, Tel-Aviv University, Tel Aviv 6997801, Israel}
\begin{abstract}
Lattice gauge theories (LGT) play a central role in modern physics,
providing insights into high-energy physics, condensed matter physics,
and quantum computation. Due to the nontrivial structure of the Hilbert
space of LGT systems, entanglement in such systems is tricky to define.
However, when one limits themselves to superselection-resolved entanglement,
that is, entanglement corresponding to specific \textcolor{black}{gauge}\textcolor{blue}{{}
}symmetry sectors (commonly denoted as superselection sectors), this
problem disappears, and the entanglement becomes well-defined. The
study of superselection-resolved entanglement is interesting in LGT
for an additional reason: when the gauge symmetry is strictly obeyed,
superselection-resolved entanglement becomes the only distillable
contribution to the entanglement. In our work, we study the behavior
of superselection-resolved entanglement in LGT systems. We employ
a tensor network construction for gauge-invariant systems as defined
by Zohar and Burrello \citep{zoharBuildingProjectedEntangled2016}
and find that, in a vast range of cases, the leading term in superselection-resolved
entanglement depends on the number of corners in the partition \textendash{}
corner-law entanglement. To our knowledge, this is the first case
of such a corner-law being observed in any lattice system.
\end{abstract}
\maketitle

\section{Introduction}

Gauge theories are central in fundamental physics, originated in continuous
field theory. In a gauge theory, the discussed system is subject to
a local (spacetime dependent) symmetry, which give rise to new fields
referred to as the gauge fields. In lattice gauge theory (LGT), \citep{wilsonConfinementQuarks1974,kogutHamiltonianFormulationWilson1975,kogutIntroductionLatticeGauge1979}
the continuous gauge theory is limited to a discretized lattice space
(or spacetime). LGTs have been a central field of study in the past
decades, leading to insights in high-energy physics (e.g., by providing
a numerical tool for the computation of the hadronic spectrum \citep{aokiFLAGReview20212022}),
condensed matter (by introducing models with \textcolor{black}{topological
order }\citep{toric}) and provides candidate models for surface codes
\citep{fowlerSurfaceCodesPractical2012,clelandIntroductionSurfaceCode2022,ioliusDecodingAlgorithmsSurface2023}.
They can be realized on quantum simulators (see, e.g., the theoretical
reviews \citep{wieseQuantumSimulatingQCD2014,zoharQuantumSimulationsLattice2015,dalmonteLatticeGaugeTheory2016,banulsReviewNovelMethods2020,banulsSimulatingLatticeGauge2020,homeierZ2LatticeGauge2021,zoharQuantumSimulationLattice2021,aidelsburgerColdAtomsMeet2021,klcoStandardModelPhysics2022,bauerQuantumSimulationHighEnergy2023,bauerQuantumSimulationFundamental2023}
and experimental demonstrations \citep{martinezRealtimeDynamicsLattice2016a,bernienProbingManybodyDynamics2017,kokailSelfverifyingVariationalQuantum2019a,schweizerFloquetApproachZ22019,milScalableRealizationLocal2020,yangObservationGaugeInvariance2020,semeghiniProbingTopologicalSpin2021,zhouThermalizationDynamicsGauge2022,riechertEngineeringLatticeGauge2022,methSimulating2DLattice2023}). 

The entanglement in LGT models is tricky to define, due to the nontriviality
of the Hilbert space \citep{casiniRemarksEntanglementEntropy2014,radicevicNotesEntanglementAbelian2014,donnellyEntanglementEntropyElectromagnetic2015,ghoshEntanglementEntropyGauge2015,huangCentralChargeEntangled2015,radicevicEntanglementWeaklyCoupled2016,donnellyGeometricEntropyEdge2016,maEntanglementCenters2016,vanacoleyenEntanglementDistillationLattice2016,pretkoEntanglementEntropyQuantum2016a,pretkoEntanglementEntropyMaxwell2018}.
However, when one restricts themselves to a single symmetry sector,
also denoted by a superselection sector in LGTs, the discussion of
entanglement becomes natural, as we explain in Sec. \ref{sec:Symmetry-resolved-entanglement-i}.
In this paper, we study symmetry resolved entanglement \citep{laflorencieSpinresolvedEntanglementSpectroscopy2014,goldsteinSymmetryResolvedEntanglementManyBody2018,PhysRevA.98.032302}
which have been extensively studied in many systems and led to interesting
discoveries \citep{sr_PhysRevB.98.041106,parez2021quasiparticle,LAFLORENCIE20161,bonsignori2019symmetry,murciano2020entanglement,Murciano_2020,Horv_th_2020,Horv_th_2021,Capizzi_2022,Castro_Alvaredo_2023,MicheleNonHermitian,Monkman_2020,Cornfeld_2019,Azses_2020,digiulio2023boundary,Calabrese_2021,murciano2023symmetryresolved,Horvath_2022,Ares_2022_MultiCharge,Feldman_2019,Lukin_2019,fraenkel2020symmetry,parez2021exact,vitale2021symmetryresolved,fraenkel2021entanglement,Scopa_2022,oblakEquipartitionEntanglementQuantum2022,sr_PhysRevB.101.235169,10.21468/SciPostPhys.10.3.054,bertini2023nonequilibrium,bertini2023dynamics,horvathChargeresolvedEntanglementPresence2023}.
The local symmetry imposes a large number of symmetry sectors, referred
to as superselection sectors. \textcolor{black}{The study of gauge
symmetry resolved (SR) entanglement is specifically interesting, as
it allows to separate the entanglement into two contributions: the
SR entanglement, which is the only distillable entanglement when the gauge symmetry is fundamental, hence has to be obeyed by any operation performed on each subsystem; and the entanglement
stemming from the division into different sectors, which detects topological
effects.}\textcolor{blue}{{} }

We use the special structure of gauge-invariant states on TN to obtain
a corner-dependent term in the superselection-resolved entanglement
and discuss the various cases in which this is the leading term, that
is, the entanglement's behavior follows a corner-law. Since studied
systems typically have a constant number of corners (e.g., rectangular
system with four corners or a strip on a cylinder with zero corners),
the superselection-resolved entanglement \textcolor{black}{in these
cases becomes independent of system size. When the harvest of entanglement
is desired as a resource for quantum technology, one may use the corner-law
to find the partition with maximal entanglement based on the number
of corners it contains.}

The rest of this paper is organized as follows: In Sec. \ref{sec:Gauge-Invariant-States-on},
we present the basics of LGTs, while focusing on pure gauge models
on a square lattice. In Sec. \ref{sec:Tensor-Network-Representation},
we present the TN construction we work with, projected entangled pair
states (PEPS), discuss its relation to entanglement, and present the
PEPS construction for gauge invariant states on which our work relies
on as proposed by Zohar and Burrello \citep{zoharBuildingProjectedEntangled2016}.
In Sec. \ref{sec:Symmetry-resolved-entanglement-i}, we discuss symmetry-resolved
entanglement and its meaning in the presence of a local gauge symmetry.
The reader who is familiar with PEPS and symmetry-resolved entanglement
may jump straight to Sec. \ref{sec:Superselection-resolved-entangle},
in which we bring all of the above together and obtain our main result
regarding the behavior of superselection-resolved entanglement in
PEPS-representable gauge-invariant states with an Abelian gauge group;
the non-Abelian case is discussed in Sec. \ref{subsec:Generalization-to-nonabelian}.
In Sec. \ref{sec:Numerical-results}, we present numerical results
on pure $\mathbb{Z}_{2}$ gauge models, which are in line with our
results and perhaps raise questions regarding the relation of entanglement
and confinement in gauge-invariant states.

An important notation clarification is required before proceeding
with the paper: Note that in the standard notation of entanglement
of two-dimensional systems, area-law (volume-law) refers the entanglement
polynomial dependence on the boundary (bulk) size of the system, while
in the standard notation of confinement (as explained in Sec. \ref{sec:Gauge-Invariant-States-on}),
area-law (perimeter-law) refers to a decay that depends on the bulk
(boundary) size of a Wilson loop. In this paper, we will use both
notations according to the discussed property and also add the bulk/boundary
notation, for clarity. 

\section{Gauge-Invariant States on a Lattice\label{sec:Gauge-Invariant-States-on}}

Gauge theories were first introduced in the context of quantum field
theory: A local (i.e., time- and space-dependent) symmetry gives rise
to so-called force-carrier fields or \emph{gauge fields}, which mediate
the interaction between matter fields \citep{peskinIntroductionQuantumField1995,altlandCondensedMatterField2010,fradkinFieldTheoriesCondensed2013}.
LGT was later introduced as a numerical discretization tool for the
study of continuous models \citep{wilsonConfinementQuarks1974,kogutHamiltonianFormulationWilson1975,kogutIntroductionLatticeGauge1979}.
However, LGTs turned out to exhibit interesting properties on their
own, and are now useful in the study of topological effects in many-body
physics and as candidates for quantum error correction codes.

\begin{figure*}[!tph]
\includegraphics[width=0.8\textwidth]{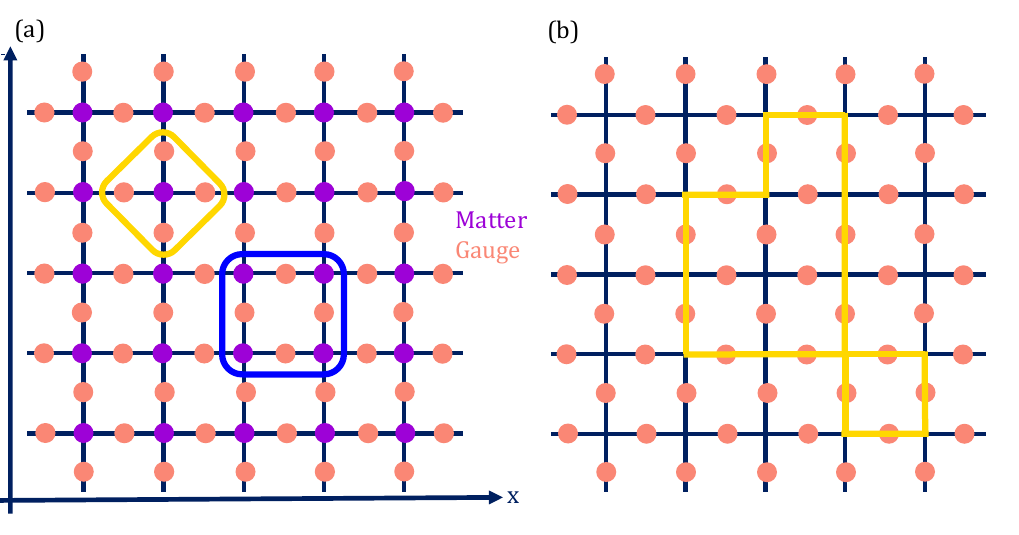}

\caption{\label{fig:lattice}The discretization of a gauge model on a lattice
graph. (a) Matter fields are placed on the lattice sites, and the
gauge fields, mitigating the interaction of matter, are placed on
the edges of the lattice. In yellow a star is denoted, that is, a
matter site and the gauge sites it interacts with, and in blue a plaquette,
that is, the minimal structure of a closed loop on the lattice. (b)
In the pure gauge case, we omit the matter sites. In this case, gauge
invariance is expressed by the requirement that the excitation on
the gauge fields may only exist on edges that form closed loops.}
\end{figure*}

As in the continuous gauge models, LGT models consist of matter degrees
of freedom (DoF) and gauge DoF. The matter DoF reside on the lattice
sites, and the gauge DoF, mediating the interaction between two lattice
sites, reside on the edges between the lattice sites, as demonstrated
in Fig. \ref{fig:lattice}a. In this paper, we focus on two-dimensional
square lattices, but the generalization could be done to higher dimensions,
nonabelian symmetries, or different lattice types. We follow the Hamiltonian
formalism of LGT in Ref. \citep{kogutHamiltonianFormulationWilson1975},
i.e., only space is discretized and time remains continuous. $G$
denotes the group corresponding to the local symmetry. For simplicity,
we focus here on Abelian symmetry groups (for the generalization to
nonabelian symmetries see, e.g., Ref \citep{zoharBuildingProjectedEntangled2016}).
A lattice site and the edges around it are denoted by a \emph{star},
as illustrated in Fig. \ref{fig:lattice}a. The gauge operator corresponding
to each star and group element $g\in G$ is defined by:
\begin{equation}
\mathcal{G}_{s}^{(g)}=\Theta_{s,\hat{x}}^{(g)}\Theta_{s,\hat{y}}^{(g)}\Theta_{s,-\hat{x}}^{(g)\dagger}\Theta_{s,-\hat{y}}^{(g)\dagger}\tilde{\Theta}_{s}^{(g)\dagger},\label{eq:gauge_op}
\end{equation}
where $\Theta_{s,\hat{e}}^{(g)},\tilde{\Theta}_{s}^{(g)}$ is the
unitary operator corresponding to $g$ applied to the gauge DoF on\textcolor{black}{{}
direction $\hat{e}$ of $s$ and the matter DoF on the lattice site
$s$, respectively. In the nonabelian case the first pair of factors
in Eq. (\ref{eq:gauge_op}) should correspond to a left group action,
while the rest should correspond to a right action. }We will focus
on pure gauge models, and the matter charges will therefore be omitted
from now on. As an example, we refer to gauge DoF with a $\mathbb{Z}_{2}$-symmetry,
that is, the edges are occupied by $1/2$-spins. The only nontrivial
element in the group corresponds to the $z$ Pauli matrix, \textcolor{black}{$\sigma^{z},$
and the gauge operator becomes
\begin{equation}
\mathcal{G}_{s}=\sigma_{s,\hat{x}}^{z}\sigma_{s,\hat{y}}^{z}\sigma_{s,-\hat{x}}^{z}\sigma_{s,-\hat{y}}^{z}.\label{eq:gauge_op_z2}
\end{equation}
}

In LGT models, the gauge operators on all of the stars in the lattice
commute with the Hamiltonian and are therefore conserved. Gauge invariance
requires that for any star $s$, in the absence of static charges,
\begin{equation}
{\color{blue}{\color{black}\mathcal{G}_{s}\k{\psi}=\k{\psi}}}.\label{eq:inv_unitary}
\end{equation}
 The requirement above imposes that the value of the gauge field going
in each node (from the left and from below) must be equal to the value
of the gauge field going out of the node (from the right and from
above). For $\mathbb{Z}_{2}$ gauge fields, this means that the number
of downward-pointing spins around each lattice site must be even. 

When $G$ is a compact Lie group, the above may be intereperted as
the Gauss law, by thinking of the nontrivial gauge charges as a flux:
the flux going in each lattice site must be equal to the flux coming
out of it, which enforces the fact that any excitation of the gauge
sites can only exist on sets of edges that construct closed loops
(see Fig. \ref{fig:lattice}b), corresponding to the continuous Gauss
law in the absence of charges, $\vec{\nabla}\cdot\vec{E}=0$. The
flux interpretation may also be valid in finite groups by introducing
modular operators as in Eq. (\ref{eq:gauge_op_z2}). In the $\mathbb{Z}_{2}$
case, one may refer to the gauge upward-pointing spins as the vacuum,
and downward-pointing spins as the flux. Downward-pointing spins may
only appear in closed loops.

As carriers of interaction, it is relevant to discuss confinement
and deconfinement in the context of models with gauge fields. In models
with matter, the confinement-deconfinement phase transition relates
to the strength of interaction between matter fields as a function
of the distance between them, that is, the length of possible gauge-field-mediated
paths between the matter particles. In pure gauge models, the above
reduces into studying loops of gauge field excitations, also known
as \emph{Wilson loops} \citep{wilsonConfinementQuarks1974}, i.e.,
operators that excite the gauge field around a closed loop of edges.
In the $\mathbb{Z}_{2}$ case, a Wilson loop is defined as
\[
\hat{W}=\otimes_{e\in W}\sigma_{e}^{x},
\]
where $e$ runs over all of the edges that the loop $W$ is composed
of and $\sigma_{e}^{x}$ is the $x$ Pauli matrix acting on the spin
on edge $e$. The expectation value of Wilson loops may decay as a
function of their area (bulk size), in which case the state is in
a confined phase, or as a function of their perimeter (boundary size),
in which case the phase is deconfined. The confined phase is a disordered
phase, and one may expect that it would imply less entanglement between
parts of the system, since large flux loops have a small amplitude,
while the deconfined phase is ordered and may exhibit larger entanglement.\textcolor{black}{{}
However, the relation of confinement and entanglement is not always
straightforward, as we demonstrate and discuss below.}

\section{Tensor Network Representation of Gauge-Invariant States\label{sec:Tensor-Network-Representation} }

The local nature of LGT models makes them suitable for representation
by means of TN ans\"{a}tze (see, e.g., \citep{krausFermionicProjectedEntangled2010,tagliacozzoEntanglementRenormalizationGauge2011,haegemanGaugingQuantumStates2015,zoharFermionicProjectedEntangled2015,zoharProjectedEntangledPair2016,dalmonteLatticeGaugeTheory2016,zoharBuildingProjectedEntangled2016,zoharCombiningTensorNetworks2018,tschirsichPhaseDiagramConformal2019,banulsReviewNovelMethods2020,emontsGaussLawMinimal2020,emontsVariationalMonteCarlo2020,montangeroLoopfreeTensorNetworks2021,banulsSimulatingLatticeGauge2020,gonzalez-cuadraRobustTopologicalOrder2020,felserTwoDimensionalQuantumLinkLattice2020,Scuchpepshard}).
Our work is based on the TN construction of gauge-invariant states
as proposed by Zohar and Burrello \citep{zoharBuildingProjectedEntangled2016}.
In this section, we present the construction for a gauge-invariant
state on a square lattice with Abelian gauge symmetry. We start by
briefly covering the TN ansatz projected entangled pair states (PEPS),
and then continue to describe PEPS states that by construction obey
gauge invariance.

\subsection{Projected Entangled Pair States (PEPS)\label{subsec:Projected-Entangled-Pair}}

\begin{figure}[!tph]
\includegraphics[width=1\linewidth]{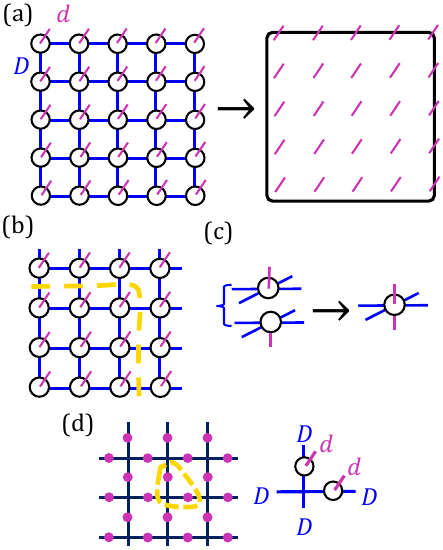}\caption{\label{fig:peps} (a) A state of an $N$-sized system with Hilbert
dimension size $d$ per site represented as a PEPS. The representation
is composed of $N$ tensors with a physical leg (pink) of dimension
$d$ and four virtual legs (blue) of dimension $D$. Contracting the
virtual legs would result in a $d^{\otimes N}$ tensor, which may
be reshaped into a $d^{N}$ vector. (b) The entanglement between two
parts of the system is upper bounded by the total dimension of the
virtual legs that cross the boundary between the two systems. (c)
A PEPO tensor may be constructed by taking two copies of the PEPS
tensor, performing a complex conjugation to one of them, and placing
them back to back. (d) A single tensor in the gauge-invariant PEPS
representation corresponds to a single lattice site and the edges
on top of it and to its right, with two physical legs of dimension
$d$ in the pure gauge case.}
\end{figure}

In this work we focus on PEPS \citep{verstraeteRenormalizationAlgorithmsQuantumMany2004,jordanClassicalSimulationInfiniteSize2008,ciracReview},
a TN ansatz used in the representation of many-body quantum states.
It is specifically relevant to states that exhibit area-law entanglement
\citep{pepsarealaw}, that is, states in which the entanglement between
two subsystems scales with the size of the boundary between them.

Consider a system of $N$ lattice sites, with Hilbert space dimension
$d$ per site. A PEPS representation of the system's state will be
composed of $N$ tensors corresponding to the $N$ lattice sites.
Each tensor has one entry of dimension $d$, denoted as \emph{the
physical leg}, representing a Hilbert space that corresponds to the
Hilbert space of a single site, and additional entries called \emph{the
virtual legs}, which connect tensors of nearest-neighbor sites. The
dimension of the Hilbert space represented by the virtual legs is
denoted by $D$ and referred to as the \emph{bond dimension}. Contracting
the virtual legs of all nearest-neighbor sites will result in a $d^{\otimes N}$
tensor representing the state of the system, which may be reshaped
into a $d^{N}$ vector, the standard representation of a quantum state.
The above is demonstrated in Fig. \ref{fig:peps}a. 

A family of algorithms relying on this construction called infinite
PEPS (iPEPS) is noteworthy: The system is assumed to be infinite and
translationally invariant, and therefore defined by a repeating PEPS
tensor. While contracting an iPEPS has been shown to be a computationally
hard problem \citep{Scuchpepshard,Haferkampspepshard}, an approximate
contraction may be done, for example, using the boundary matrix product
states method \citep{BMPS} or the corner transfer matrix method \citep{ctmrgNishino,ctmrgOrus2009,ctmrgOrus2012},
allowing for the computation of expectation values and other properties
of the state represented by an iPEPS. The PEPS and iPEPS ans\"{a}tze
are central numerical tools in the study of strongly correlated two-dimensional
systems, and has been used for finding ground states \citep{CorbozHubbard,CorboztJ,Xiangkhaf,PicotPRBB2016,KshetrimayumXXZ,kshetrimayumTensorNetworkInvestigation2020,boosCompetitionIntermediatePlaquette2019},
thermal states \citep{CzarnikfinT2012,CzarnikfinT2015,KshetrimayumfinT2019,CzarniKitaevfinT,CzarnikSSlandfinT,Mondal2020}
and non-equilibrium steady-states \citep{KshetrimayumNatcomm2017,Czarnikevolution2019,Hubig2019,Kshetrimayum2DMBL,Kshetrimayum2DTC,Dziarmaga,kaneko2021tensornetwork}
in two spatial dimensions. 

When discussing density matrices rather than states, as is often the
case in the study of entanglement, it is useful to define projected
entangled pair operators (PEPO), and the operator version of iPEPS,
iPEPO. A PEPO node has two physical indices rather than one \textemdash{}
corresponding to two physical entries of the density matrix. A PEPS
tensor may be used to create a PEPO tensor analogously to the definition
of a density matrix out of a state vector, $\rho=\k{\psi}\b{\psi}$.
The PEPO tensor is constructed by taking two copies of the tensor,
performing a complex conjugation to one of them, and placing them
back to back, as demonstrated in Fig. \ref{fig:peps}c. 

\subsubsection{Entanglement in Tensor Network States}

We start by defining the entanglement measures that will be referred
to in this work: For a given system in a pure state $\k{\psi}$, composed
of two subsystems $A,B$, the reduced density matrix (RDM) of $A$
is defined to be
\begin{equation}
\rho_{A}=\Tr_{B}\left[\k{\psi}\b{\psi}\right],\label{eq:rdm}
\end{equation}
where $\Tr_{B}$ denotes the tracing out of the degrees of freedom
of subsystem $B$. The entanglement between subsystems $A$ and $B$
is standardly quantified by the von Neumann (vN) entropy \citep{Wilde}:
\begin{equation}
S(\rho_{A})=-\Tr\left[\rho_{A}\log\rho_{A}\right].\label{eq:vN}
\end{equation}
Complementary entanglement measures are the R\'enyi entropies \citep{Wilde}:
\begin{equation}
S^{(n)}(\rho_{A})=\frac{1}{1-n}\log\Tr\left[\rho_{A}^{n}\right].\label{eq:renyi}
\end{equation}
Note that $\lim_{n\rightarrow1}S^{(n)}(\rho_{A})=S(\rho_{A}).$ While
the R\'enyi entropies are entanglement monotones, they do not possess
some useful properties of the vN entropy \citep{Wilde}. However,
the fact that their computation does not require taking a log of the
matrix but only multiplying $n$ copies of it, makes it more accessible
numerically and experimentally \citep{pastMeas_2012_entropy,Daley,Pichler,IslamGreiner,PichlerPRX,pastMeas_apr_2019_experimental_purity,pastMeas_feb_2018_entropy_spin,pastMeas_feb_2018_entropy_spin_hubbard_reference,pastMeas_may_2019_local_entropy,CGSMeas_jun_2019,pastMeas_jul_2020_entropy_experiment,pastMeas_jun_2020,feldmanEntanglementEstimationTensor2022},
and sometimes also useful analytically by means of the replica trick,
as can be seen, e.g., in Refs. \citep{goldsteinSymmetryResolvedEntanglementManyBody2018,PhysRevA.98.032302,knauteEntanglementConfinementLattice2024}
and in Sec. \ref{subsec:Transfer-matrix-method} below.

We study the relation of the bond dimension $D$ to the entanglement
between two subsystems $A,B$ of the system: All quantum states may
be written in their Schmidt decomposition \citep{SCHOLLWOCK_2011}:
\begin{equation}
\k{\psi}_{AB}=\sum_{i=1}^{\tilde{D}}\psi_{i}\k i_{A}\k i_{B},\label{eq:schmidt}
\end{equation}
where $\left\{ \k i_{A}\right\} ,\left\{ \k i_{B}\right\} $ are orthonormal
sets of states in the Hilbert space of $A,B$, respectively. From
this decomposition, it straightforwardly follows that $\rho_{A},\rho_{B}$
have the same eigenvalues, $\left\{ \left|\psi_{i}\right|^{2}\right\} _{i=1}^{\tilde{D}}$.
$\tilde{D}$ is thus the number of nonzero eigenvalues (rank) of the
RDMs. Any TN state may be brought to a form in which each set of bond
indices residing on the boundary between subsystems $A$ and $B$
corresponds to a single coefficient $\psi_{i}$ \citep{SCHOLLWOCK_2011},
often referred to as the orthogonal or isometric form \citep{zaletelIsometricTensorNetwork2020}.
In a PEPS, $\tilde{D}$ is bounded from above by the dimension of
all of the virtual legs crossing the boundary between $A$ and $B$,
$\tilde{D}\le D^{\left|\text{boundary}\right|}$, as illustrated in
Fig. \ref{fig:peps}b. Therefore, it is clear that the requirement
that D is constant (as in translationally-invariant representations
such as iPEPS) or efficiently dependent on the system size (as in
all efficiently-representable PEPS states) implies that the state's
entanglement must obey an area-law for any partition $A,B$. 

\subsection{Gauge-Invariant PEPS\label{subsec:Gauge-Invariant-PEPS}}

In this section we briefly review the construction of gauge-invariant
PEPS, and focus on Abelian gauge symmetries. For a detailed review,
including the generalization to nonabelian symmetries, see Ref. \citep{zoharBuildingProjectedEntangled2016}.
Based on the interpretation to the Gauge field excitations as flux,
the PEPS tensor is chosen to represent a single lattice site and the
edges coming out of it to the top and to the right (see Eq. (\ref{eq:allowed_terms})):
The flux going into the node from the bottom and left must be equal
to the flux going out of the node from the top and right. This may
be imposed by assigning the Hilbert space corresponding to the virtual
legs that has the same symmetry as the one on the physical gauge sites,
as detailed below.

The local operators applied to each gauge link are elements of some
finite or compact symmetry group $G$. The unitary operator corresponding
to the group element $g\in G$ is denoted by $\Theta^{(g)}$, and
obeys
\[
\Theta^{(g)}\k j=e^{i\phi_{j}^{(g)}}\k j,
\]
where $j$ corresponds to states in the Hilbert space of a single
gauge link (group representations) and $\phi_{j}^{(g)}$ is the acquired
phase. 

We utilize the construction above and span the Hilbert space on the
virtual legs using irreps of the same group $G$. The Hilbert space
corresponding to the virtual legs may be of different dimension, but
the symmetry of $G$ must be obeyed. In this way, the flux value on
the virtual legs becomes equivalent to the virtual flux value on the
physical gauge legs. As in the physical Hilbert space, we define the
unitary operator corresponding to the group element $g\in G$ on the
bond Hilbert space to be $\theta^{(g)}$:
\[
\theta^{(g)}\k j=e^{i\phi_{j}^{(g)}}\k j.
\]

We now impose the local symmetry, as required in Eq. (\ref{eq:inv_unitary}).
The gauge operators apply to the physical legs of three neighboring
PEPS tensors:

\begin{equation}\label{eq:gauge_inv}
\adjustbox{valign=m}{\includegraphics[width=0.35\linewidth]{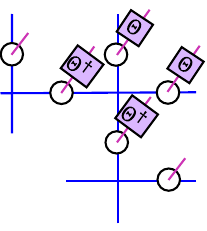}}=
\adjustbox{valign=m}
{\includegraphics[width=0.35\linewidth]{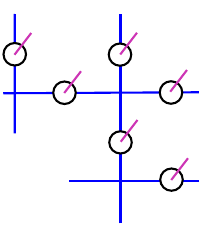}}
\end{equation}

The requirement in Eq. (\ref{eq:gauge_inv}) may be translated into
a connection between the physical and virtual Hilbert spaces:

\begin{equation}\label{eq:gauge_inv_node}
\begin{aligned}
\adjustbox{valign=m}{\includegraphics[width=0.86\linewidth]{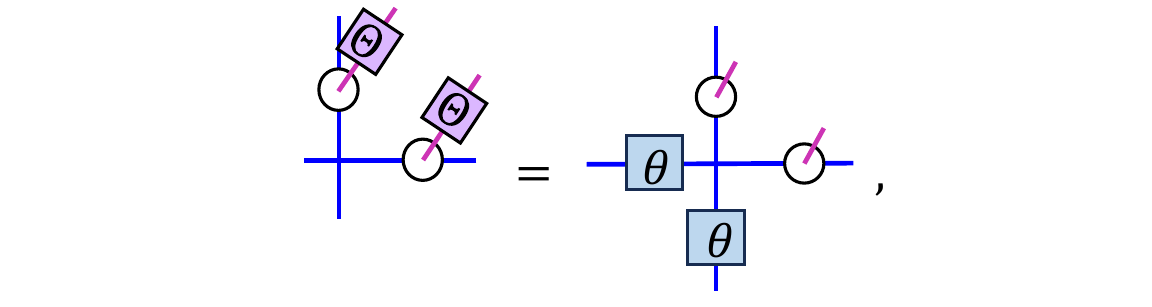}} \\
\adjustbox{valign=m}{\includegraphics[width=0.43\linewidth]{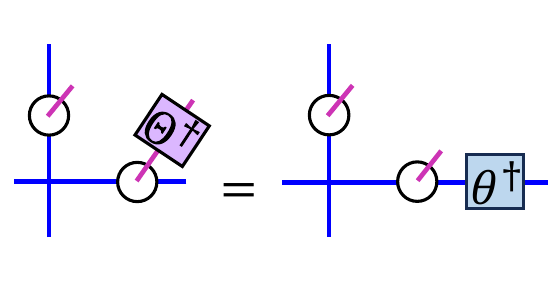}},
\adjustbox{valign=m}
{\includegraphics[width=0.365\linewidth]{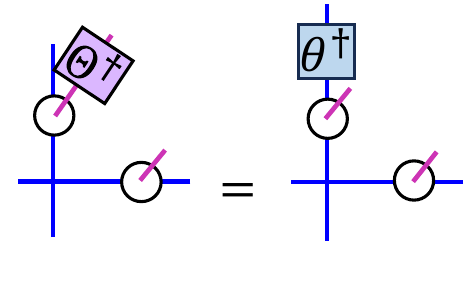}}.
\end{aligned}
\end{equation}In order to impose Eq. (\ref{eq:gauge_inv_node}), we require that
the only nonzero terms in the gauge invariant PEPS tensors are the
ones in which the outgoing flux on the top and right physical legs
equals the outgoing flux on the top and right virtual legs, respectively.
The total flux on the outgoing top and right virtual legs is required
to be equal to the total flux on the ingoing bottom and left virtual
legs:\begin{equation}\label{eq:allowed_terms}
\textcolor{purple}{
\begin{aligned}
& \adjustbox{valign=m}
{\includegraphics[width=0.24\linewidth]{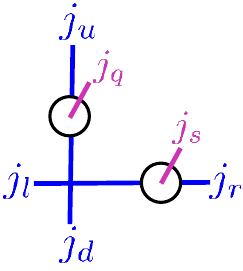}},\\&
j_q \doteq j_u,
\\&
j_s \doteq j_r,
\\&
j_u + j_r = j_l + j_d
,
\end{aligned}
}
\end{equation} where $\doteq$ stands for equivalence of the physical and virtual
group elements.

The states represented by PEPS that obey the requirement above are
gauge-invariant by construction. Since the values of the physical
indices are repeated on the upper and right virtual legs, we omit
the drawing of the physical indices below for simplicity. Ref. \citep{emontsFindingGroundState2023}
relies on the construction above to find PEPS representations to the
ground state of the toric code model with added local magnetic field,
and obtain an ansatz that requires $D=4$ for the confined phase and
$D=2$ for the deconfined phase, as one might expect, recalling that
the larger the bond dimension, the larger the entanglement may be.
Ref. \citep{zoharWilsonLoopsArea2021} describes a minimal model for
$\mathbb{Z}_{2}$ gauge DoF with rotational invariance with $D=2$:

\begin{equation}\label{eq:zohar_model}
\begin{aligned}
&\adjustbox{valign=m}{\includegraphics[width=0.2\linewidth]{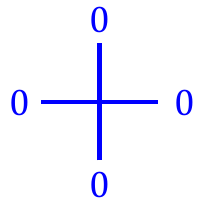}} = \alpha, \\&
\adjustbox{valign=m}{\includegraphics[width=0.18\linewidth]{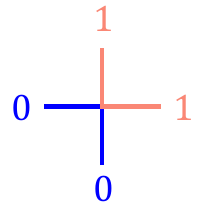}}=
\adjustbox{valign=m}
{\includegraphics[width=0.18\linewidth]{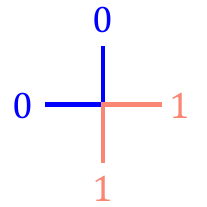}} = 
\adjustbox{valign=m}
{\includegraphics[width=0.18\linewidth]{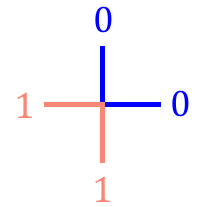}} 
=
\adjustbox{valign=m}
{\includegraphics[width=0.18\linewidth]{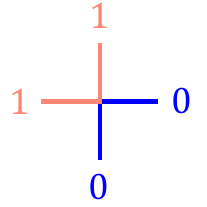}}=\beta, \\&
\adjustbox{valign=m}
{\includegraphics[width=0.18\linewidth]{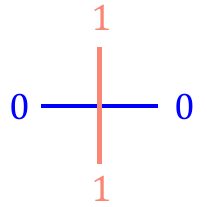}} = 
\adjustbox{valign=m}
{\includegraphics[width=0.18\linewidth]{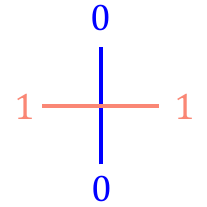}} 
=\gamma,
\adjustbox{valign=m}
{\includegraphics[width=0.18\linewidth]{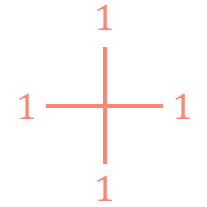}}=\delta,
\end{aligned}
\end{equation}and analyze it analytically for several cases. Specifically, when
$\beta\ll\delta\lesssim\alpha,$ the PEPS is shown to represent a
confined state if $\gamma=0$, where $\gamma>0$ results in a deconfined
state.

\textcolor{black}{One may also construct PEPS-representable states
with an infinite symmetry gauge group, such as U(1), by truncating
the link Hilbert space in the representation basis, and keeping only
a finite number of representations \citep{zoharFormulationLatticeGauge2015}
(for example, in U(1) where representations corresponds to the angular
momentum value along the group circle, one may restrict the link Hilbert
space to angular momenta $-m\dots m$, resulting in $d=2m+1$). The
physical Hilbert space dimension $d$ then becomes finite, and therefore
the bond dimension may also be finite and the state representable
by PEPS.}

\section{Symmetry-resolved entanglement in gauge-invariant states\label{sec:Symmetry-resolved-entanglement-i}}

Before we dive into the many definitions in this section, it is noteworthy
that in gauge-invariant theories, the structure of the Hilbert space
is different from the general case: In a general system, the Hilbert
space of systems $A,B$ is of the structure $\mathcal{H}_{AB}=\mathcal{H}_{A}\otimes\mathcal{H}_{B}.$
However, once gauge-noninvariant states are excluded from $\mathcal{H}_{AB}$,
the Hilbert space loses its tensor-product form, and the Hilbert spaces
$\mathcal{H}_{A},\mathcal{H}_{B}$ become ill-defined, and in turn
also $\rho_{A},\rho_{B}$ and the entanglement measures which rely
on them. In this paper, we follow the standard method for solving
this problem \citep{buividovichEntanglementEntropyGauge2008,donnellyDecompositionEntanglementEntropy2012,ghoshEntanglementEntropyGauge2015,aokiDefinitionEntanglementEntropy2015,soniAspectsEntanglementEntropy2016}
by embedding the gauge-invariant Hilbert space in the full Hilbert
space $\mathcal{H}_{A}\otimes\mathcal{H}_{B}$. As we will shortly
explain, when defining symmetry-resolved entanglement and limiting
the Hilbert space into a single block in $\rho_{A}$ and its corresponding
block in $\rho_{B}$, the problem above fades away: The gauge-invariant
Hilbert subspace is naturally in a product form, and the embedding
becomes trivial. We will continue concentrating on the Abelian case,
and will briefly consider the nonabelian case in Sec. \ref{subsec:Generalization-to-nonabelian}.

Symmetry-resolved entanglement measures, as defined in Refs. \citep{laflorencieSpinresolvedEntanglementSpectroscopy2014,goldsteinSymmetryResolvedEntanglementManyBody2018},
study the contribution to entanglement from different symmetry sectors
and point to relations between entanglement and symmetry. The behavior
of such measures in systems with global symmetries has been heavily
studied in recent years, both analytically and numerically \citep{sr_PhysRevB.98.041106,parez2021quasiparticle,LAFLORENCIE20161,bonsignori2019symmetry,murciano2020entanglement,Murciano_2020,Horv_th_2020,Horv_th_2021,Capizzi_2022,Castro_Alvaredo_2023,MicheleNonHermitian,Monkman_2020,Cornfeld_2019,Azses_2020,digiulio2023boundary,Calabrese_2021,murciano2023symmetryresolved,Horvath_2022,Ares_2022_MultiCharge,Feldman_2019,fraenkel2020symmetry,parez2021exact,vitale2021symmetryresolved,fraenkel2021entanglement,Scopa_2022,oblakEquipartitionEntanglementQuantum2022,sr_PhysRevB.101.235169,10.21468/SciPostPhys.10.3.054,bertini2023nonequilibrium,bertini2023dynamics,horvathChargeresolvedEntanglementPresence2023},
and as a principle for experimental measurement protocols \citep{CGSMeas_jun_2019,Lukin_2019}.
The study of symmetry-resolved entanglement may point to instances
of dissipation in open systems dynamics \citep{vitale2021symmetryresolved}
or effects such as topological phase transitions \citep{Cornfeld_2019,fraenkel2020symmetry,Azses_2020}. 

\begin{figure}[th]
\includegraphics[width=1\linewidth]{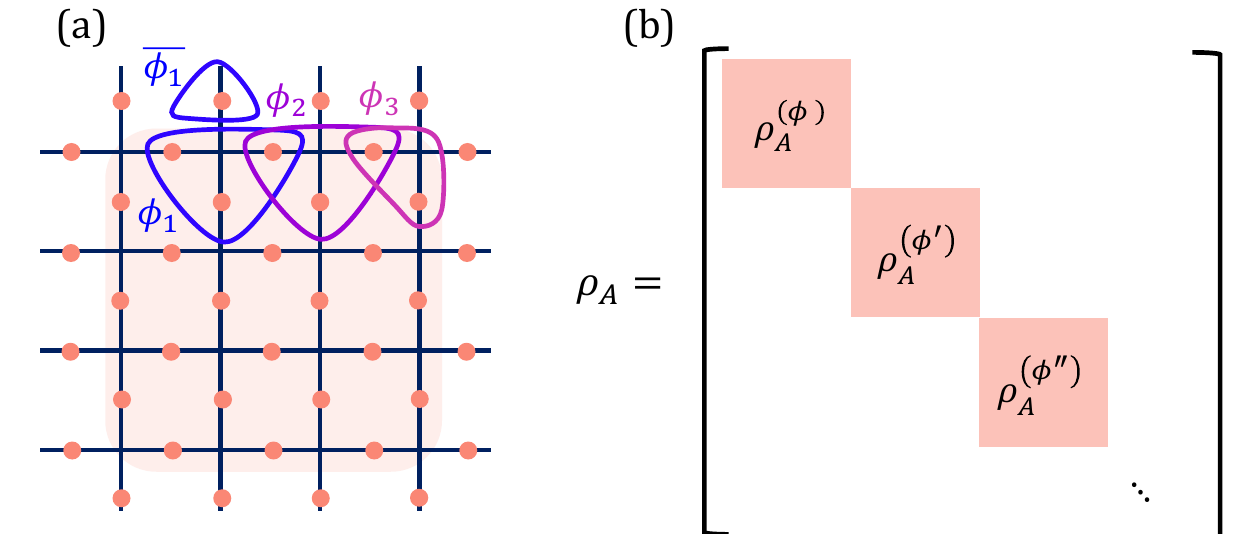}

\caption{\label{fig:block_diag}(a) The superselection sectors are subspaces
of the Hilbert space of subsystem $A$, defined by the flux on the
star-parts on the boundary between the subsystem and its environment.
In circles are examples of star parts on the boundary between the
subsystems. (b) The RDM is deconposed of blocks, each block corresponding
to a single superselection sector.}
\end{figure}

\textcolor{black}{First, we show that thanks to the Gauss law, the
RDM $\rho_{A}$ is block diagonal, with different blocks corresponding
to different values of the flux going through the boundary between
subsystems $A,B,$ as illustrated in Fig. \ref{fig:block_diag}b.
We work in the eigenbasis of the unitaries $\Theta_{g}$, $\left\{ \k j\right\} _{j=1}^{d}$,
and write the state of the full system as
\[
\k{\psi}=\sum_{\vec{j}}\psi_{\vec{j}}\k{\vec{j}},
\]
where $\left[\vec{j}\right]_{e}$ is the state on the site $e$. For
each star $s$ on the boundary of subsystems $A,B$, we denote by
$s_{A},s_{B}$ the star parts residing on subsystems $A,B$, so that
$s=s_{A}s_{B}$. From Eq. (\ref{eq:inv_unitary}) it is required that
the value of the total flux on $s_{A}$ fully determines the value
of the total flux on $s_{B}$. We denote by $\phi_{s},\overline{\phi_{s}}$
the total flux values on the star parts $s_{A},s_{B}$ across the
boundary in subsystems $A,B$, respectively, and by $\left\{ \vec{j}\right\} ^{\phi_{s}},\left\{ \vec{j}\right\} ^{\overline{\phi_{s}}}$
the subspaces of $A,B$ that correspond to the flux $\phi_{s}$. The
state may then be written as
\[
\k{\psi}=\sum_{\phi_{s}}\sum_{\vec{j}\in\left\{ \vec{j}\right\} ^{\phi_{s}},\vec{j}^{\prime}\in\left\{ \vec{j}\right\} ^{\overline{\phi_{s}}}}\psi_{\vec{j}\oplus\vec{j}^{\prime}}\k{\vec{j}}_{A}\k{\vec{j}^{\prime}}_{B}.
\]
Tracing out the DoF of subsystem $B$, and noting that states corresponding
to different fluxes are orthogonal, the remaining RDM is
\[
\rho_{A}=\sum_{\phi_{s}}\sum_{\vec{j}\in\left\{ \vec{j}\right\} ^{\phi_{s}},\vec{\tilde{j}}\in\left\{ \vec{j}\right\} ^{\phi_{s}}}\left[\sum_{j^{\prime}\in\left\{ \vec{j}\right\} ^{\overline{\phi_{s}}}}\psi_{\vec{j}\oplus\vec{j}^{\prime}}\psi_{\vec{\tilde{j}}\oplus\vec{j}^{\prime}}^{*}\right]\k{\vec{j}}_{A}\b{\vec{\tilde{j}}}_{A}.
\]
The obtained RDM is block diagonal, each block corresponding to a
different value of $\phi_{s}$. One may repeat this argument for all
stars on the boundary, and obtain a block diagonal RDM where each
block corresponds to the set of flux values $\left\{ \phi_{s}\right\} _{s\in\text{boundary}[AB]}$,
which for brevity we will henceforth denote by $\phi$. We denote
by $\rho_{A}^{(\phi)}$ the block corresponding to the flux combination
$\phi$, and note that $\rho_{A}=\oplus_{\phi}\rho_{A}^{\phi}$. }

One may now define the symmetry-resolved entanglement as the entanglement
obtained from each block separately: the symmetry-resolved vN and
R\'enyi entropies are defined in analogy with Eqs. (\ref{eq:vN},\ref{eq:renyi}),
respectively:\textcolor{black}{
\begin{equation}
S_{A}(\phi)=-\Tr\left(\rho_{A}^{(\phi)}\log\rho_{A}^{(\phi)}\right),\label{eq:symresolved}
\end{equation}
\begin{equation}
S_{A}^{(n)}(\phi)=\frac{1}{1-n}\log\Tr\left(\left[\rho_{A}^{(\phi)}\right]^{n}\right).\label{eq:symresolved_renyi}
\end{equation}
}Note that the symmetry-resolved vN and \textcolor{black}{the moments
of the }R\'enyi entropies sum to their nonresolved analogs, $\sum_{\phi}S_{A}(\phi)=S_{A},\sum_{\phi}{\color{black}\exp}\left({\color{black}\left(1-n\right)}S_{A}(\phi)^{(n)}\right)=\exp\left(\left(1-n\right)S_{A}^{(n)}\right)$.
It is also interesting to define the normalized symmetry-resolved
entanglement measures, obtained by normalizing the blocks, 
\begin{align}
\overline{\rho_{A}^{(\phi)}} & =\rho_{A}^{(\phi)}/\Tr\rho_{A}^{(\phi)},\nonumber \\
\overline{S_{A}}(\phi) & =-\Tr\left(\overline{\rho_{A}^{(\phi)}}\log\overline{\rho_{A}^{(\phi)}}\right),\label{eq:resolved_vn}\\
\overline{S_{A}^{(n)}}(\phi) & =\frac{1}{1-n}\log\Tr\left(\overline{\rho_{A}^{(\phi)}}^{n}\right).\label{eq:resolved_renyi}
\end{align}
We follow common notation and denote the different blocks $\rho_{A}^{\left(\phi\right)}$
by \emph{superselection sectors}, and the entanglement measures $\overline{S_{A}}\left(\phi\right),\overline{S_{A}^{(n)}}\left(\phi\right)$
by \emph{superselection-resolved }(SR)\emph{ entanglement}. In the
rest of this paper, we follow the normalized definition in Eqs. (\ref{eq:resolved_vn},\ref{eq:resolved_renyi}),
as justified later in this section. The case in which the symmetry-resolved
entanglement is independent of the block is called \emph{equipartition
}\citep{sr_PhysRevB.98.041106}\emph{. }It is observed analytically
and numerically in many systems with one and two space dimensions,
although some models without equipartition have been observed in 1+1D
systems \citep{vitale2021symmetryresolved,sr_PhysRevB.103.L041104,zhaoChargedMomentsW32022,horvathChargeresolvedEntanglementPresence2023,folignoEntanglementResolutionFree2023,ghasemiUniversalThermalCorrections2023,kusukiSymmetryresolvedEntanglementEntropy2023}.

We now note an important observation: The contribution to the full
(nonresolved) entanglement comes from two sources: the division of
$\rho_{A}$ into blocks and the contribution of each block (SR entanglement)\textcolor{black}{.
For example, for the von Neumann entropy one has
\begin{align}
S_{A} & =\sum_{\phi}p(\phi)\overline{S_{A}}(\phi)-\sum_{\phi}p(\phi)\log p(\phi),\label{eq:full_resolved_cont}
\end{align}
where $p(\phi)=\Tr\left(\rho_{A}^{(\phi)}\right).$ The off-diagonal
terms (coherence) in $\rho_{A}$ is limited to the superselection
sector blocks, that is, relevant to the first term only. }If one limits
themselves only to gauge-conserving operators and observables on the
stars and star-parts in subsystem $A$ (whether equipartition applies
or not), the contribution of $\rho_{A}$'s division into blocks becomes
inaccessible: The effect of the division into blocks on any observed
quantity can be simulated classically. This makes the normalized SR
entanglement a measure for the accessible entanglement, sometimes
referred to as distillable entanglement. The limitation to gauge-conserving
operators applies in simulated high-energy models, and may also be
applied in condensed matter models, further motivating the study of
SR entanglement in such gauge-invariant states.

\subsection{Superselection-resolved entanglement in gauge invariant tensor network
states\label{sec:Superselection-resolved-entangle}}

In this section we derive our main result, characterizing the behavior
of superselection-resolved entanglement in gauge-invariant lattice
states. As in Sec. \ref{sec:Tensor-Network-Representation}, we focus
on a pure gauge model on a square lattice obeying an Abelian symmetry;
the nonabelian case will considered in Sec. \ref{subsec:Generalization-to-nonabelian}.
The generalization to different lattice types, higher dimensions,
or gauge models with matter is straightforward. 

\begin{figure}[!tph]
\includegraphics[width=1\linewidth]{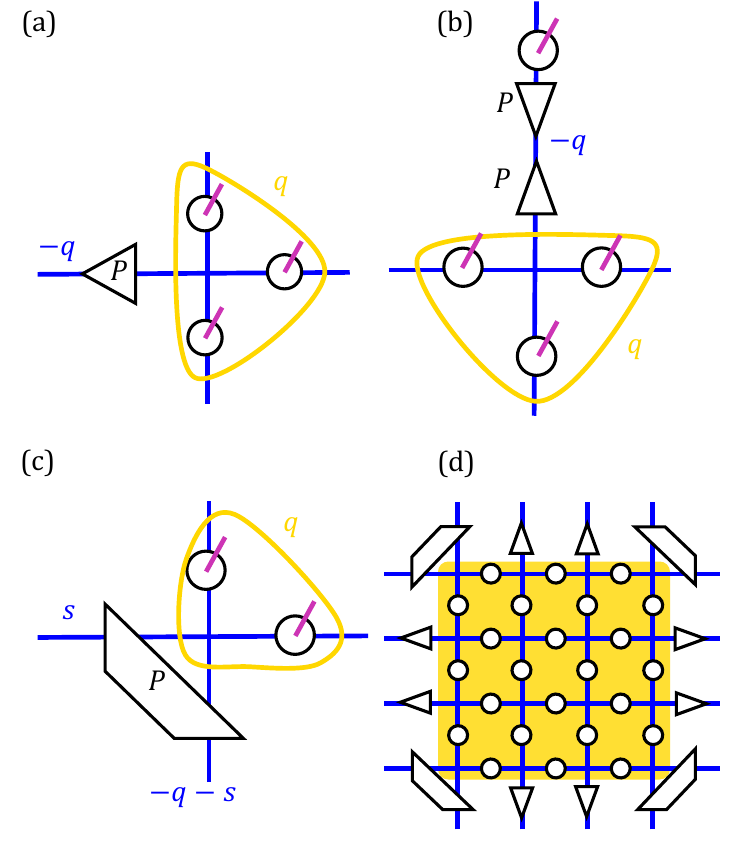}

\caption{\label{fig:peps_projection}The projection onto specific \textcolor{blue}{flux}
of a star part may be done by projecting the virtual legs going into
or out of the star part. (a) When the completing site of the star
belongs to a different PEPS site than the sites in the star part,
the projection may be done on one of the ingoing virtual legs. The
physical charge on the star part is denoted by $q$. Since the total
charge on the star must equal 0, the charge on the bond must equal
the bond-representation equivalent of $-q$. (b) When the completing
site of the star belongs to the same PEPS node as one of the sites
on the star, the node may be split into two tensors, and the projection
applied to the new bond dimension. (c) In the case of a two-site star
part (corner), the projection is done on two virtual legs together
and imposes the total charge on both of them, rather than the chrage
of each of them separately. (d) The projection onto a single superselection
block is done by projecting all star parts. }
\end{figure}

Our analysis is based on the relation of entanglement and bond dimension
described in Sec. \ref{subsec:Projected-Entangled-Pair}. This relation
has two implications: (a) The projection to a specific superselection
sector, that is, a specific set of RDM eigenvalues, may be obtained
solely by applying a projection to the virtual Hilbert subspace, as
illustrated in Fig. \ref{fig:peps_projection}. Such a projection
becomes natural for PEPS constructed as in the protocol described
in Sec. \ref{subsec:Gauge-Invariant-PEPS}: A projection onto a specific
physical charge turns into a projection onto the corresponding charge
on the virtual legs. If a star-part contains only one of the lattice
sites in a PEPS tensor, the decomposition of the tensor is required,
as depicted in Fig. \ref{fig:peps_projection}b. (b) After the projection
onto a superselection sector on the boundary, the dimension product
of the virtual indices \textcolor{black}{bounds} the number of nonzero
eigenvalues of $\rho_{A}\left(\phi\right),\rho_{B}\left(\phi\right),$
which in turn bound the SR entanglement from above. This understanding
is the basis of our results.

The projection onto a single block is obtained by applying many local
projections, on all of the star-parts on the system, see Fig. \ref{fig:peps_projection}d.
Note that the \textcolor{black}{partition boundary divides the stars
into subgroups of three and one gauge sites (Fig \ref{fig:peps_projection}a,b).
Determining the flux on a star-part on the edge therefore fully determines
the flux on the single-site star-part. On the corners of the partition,
the star-parts on both subsystems contain two gauge sites. Determining
the flux on one of the star-parts therefore does not determine the
flux of any single gauge site, and therefore also does not determine
the cha}rge on any single virtual leg (Fig. \ref{fig:peps_projection}c).

We refer to the case where $D=d$. This requirement is obeyed in PEPS
representations of various models, such as the toric code model with
string tension, displaying a topological-polarized phase transition
\citep{orusTopologicalTransitionsMultipartite2014}, thermal state
inspired TN states undergoing a symmetry enriched topological phase
transition\citep{liuTopologicalQuantumPhase2023}, and the TN construction
proposed in Ref. \citep{zoharWilsonLoopsArea2021} and presented in
Eq. (\ref{eq:zohar_model}), displaying both confined and deconfined
phases. We recall that making a projection of a single site into a
specific charge is equivalent to projecting the virtual leg next to
it onto the corresponding charge. In the case of $D=d,$ this results
in a projection of the virtual leg on the boundary between the subsystems
onto a Hilbert space of size 1. Effectively, the projected tensor
has bond dimension 1 on all of the edges on the subsystems' boundary,
adding no contribution to the bond dimension product and in turn,
to the SR entanglement. The only exception are the corners of the
system, which may require a bond dimension $D=d$ on the corner. We
then obtain a dependence of the entanglement on the number of corners
on the boundary between subsystems $A$ and $B$ \textendash{} a corner-law
entanglement:\textcolor{black}{
\begin{equation}
\overline{S}(\phi)=O\left(\log\left[\#\text{corners}(A,B)\right]\right),\label{eq:corner-law-1}
\end{equation}
}which to our knowledge, has not been observed in any other system
before. 

\textcolor{black}{We note that one may choose the bipartition around
the corners such that the star part sizes are odd and the corner contribution
vanishes, as can be seen in Sec. \ref{subsec:Numerical-results} below,
but this bipartition may be considered less natural as the one in
Fig. \ref{fig:peps_projection}d.}

We stress again the point made in Sec. \ref{sec:Symmetry-resolved-entanglement-i}:
If one restricts themselves to applying and measuring only gauge-invariant
operators, as is often the case, the only quantum correlation between
the systems is the SR entanglement. The correlations between two parts
of the system stemming from the division into superselection sectors
is completely classical, in the sense that it can be simulated classically
and may not be harvested, for example, to create Bell pairs. Our result
shows that for all states in which $D=d$, the SR entanglement is
constant and small for typical partitions (e.g., rectangular systems),
and the quantum correlation between the subsystems in this case is
thus strictly limited.

In the general case where $D>d$, the projection in Fig. \ref{fig:peps_projection}
results in virtual legs of dimension $D/d$. The SR entanglement then
obeys an area law, proportionate to the number of virtual legs on
the boundary between subsystems $A$ and $B$:\textcolor{black}{
\begin{equation}
{\color{blue}{\color{black}\overline{S}\left(\phi\right)=O\left(\log\left[\text{Area}\left(A,B\right)\right]\right).}}\label{eq:area-law}
\end{equation}
} One may recall again the result of Emonts et al. \citep{emontsFindingGroundState2023}:
A deconfined phase, intuitively believed to have larger entanglement,
typically requires $D=2d,$ where the confined phase may be represented
to a good approximation using $D=d$. The relation between entanglement,
whether SR or not, and confinement, is not always intuitive as is
the case in Ref. \citep{emontsFindingGroundState2023}. For example,
the model in Eq. (\ref{eq:zohar_model}) is constructed to have $D=d$,
but may represent both confined and deconfined phases. In appendix
\ref{sec:appendix-Superselection-resolved-entangle-1} we present
the ``perfectly confined'' state in $d=2$, i.e., a state for which
a Wilson loops' expectation value on its area is exactly exponentially-dependent
on its area. Such a state requires $D=4$, but as we show in appendix
\ref{sec:appendix-Superselection-resolved-entangle-1}, its SR entanglement
is constant \textendash{} the entanglement's dependence on system's
properties is trivial and not even corner-law.\textcolor{black}{{} Ref.
\citep{xuEntanglementGaugeTheories2023} shows that in an isometric
mapping between Kitaev's toric code model and a LGT model, both presenting
a confined-deconfined transition, the entanglement spectrum changes,
indicating that entanglement in gauge-invariant models indeed behaves
differently in LGT than in general models. }From the above, we understand
that the relation of entanglement and confinement in LGT models requires
further study.

\subsection{\textcolor{black}{Generalization to nonabelian theories\label{subsec:Generalization-to-nonabelian}}}

\textcolor{black}{In nonabelian theories, the requirement (\ref{eq:gauge_inv})
becomes slightly more complex and therefore also the PEPS representation
of gauge-invariant states. However, the idea behind our argument remains
the same, and a corner law is obtained in the representation with
minimal bond dimension, as we now explain.}

\textcolor{black}{In nonabelian symmetries, the gauge requirement
Eq. (\ref{eq:inv_unitary}) becomes a requirement that all stars are
singlets of the gauge symmetry group. The states of each physical
leg are then of the form $\k{jmm^{\prime}},$ where $j$ stands for
the representation of the group and $m,m^{\prime}$ are states in
the representation $j$, corresponding to left- and right-applied
operators (for full details, see Ref. \citep{zoharBuildingProjectedEntangled2016}).
The bond states are then of the form $\k{\tilde{j}\tilde{m}},$ where
$\tilde{j}$ is a group representation and $\tilde{m}$ an element
in $\tilde{j}$. }

\textcolor{black}{To form a singlet, the two parts of a star should
correspond to some representation $j$ and its conjugate, and the
values of $m$ should match as well. The density matrix of $A$ then
breaks into blocks corresponding to specifying $j$ and $m$ for all
the stars cut by the boundary of $A$. As a result, when the bond
dimension is minimal, so that each representation in the physical
space appears once in the virtual space (in which case the virtual
bond dimension is smaller than the physical one, since the virtual
bonds carry a single index $\tilde{m}$, while the physical bonds
carry two indices $m$, $m^{\prime}$), for each block all the virtual
legs at the boundary are fixed except for the corners, leading again
to a corner law. }

\textcolor{black}{We also note that, for each representation $j$
and its conjugate the amplitude of the different states $m$ in the
singlet state is equal in magnitude, leading to equipartition between
blocks with the same $j$ value \citep{soniAspectsEntanglementEntropy2016,vanacoleyenEntanglementDistillationLattice2016};
equipartition between blocks with different $j$ values is not guaranteed
by symmetry, as in the Abelian case. Finally, As in the Abelian case,
gauge models with infinite symmetry groups such as SU(2) may be truncated
keeping a finite number of representations per link, such that they
are representable by PEPS and the argument applies to them as well.}

\section{Numerical results\label{sec:Numerical-results}}

\subsection{Transfer matrix method for system-size dependent properties\label{subsec:Transfer-matrix-method}}

In this section, we go over the technique we used for computing the
confinement measure, i.e., area- (bulk-) law dependence of Wilson
loop expectation values, following Ref. \citep{zoharWilsonLoopsArea2021}.
Using the same line of thought, we follow Ref. \citep{knauteEntanglementConfinementLattice2024}
and describe the method we used for computing the area- (boundary-)
law dependence of full and SR R\'enyi entropies.

Wilson loop expectation values may be computed by a tiling of PEPS
tensors. We denote traced PEPO nodes, i.e., PEPO nodes with contracted
physical indices, as follows:

\begin{equation}\label{eq:traced}
\begin{aligned}
& \adjustbox{valign=m}
{\includegraphics[width=0.2\linewidth]{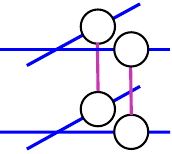}}
&=&\quad
\adjustbox{valign=m}
{\includegraphics[width=0.1\linewidth]{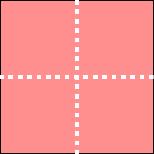}}
\text{ },
\\
&\adjustbox{valign=m}
{\includegraphics[width=0.2\linewidth]{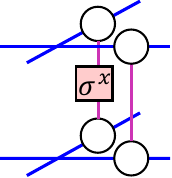}}
&=&\quad
\adjustbox{valign=m}
{\includegraphics[width=0.1\linewidth]{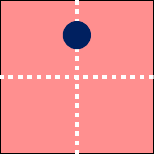}}
\text{ },
\\
&\adjustbox{valign=m}
{\includegraphics[width=0.2\linewidth]{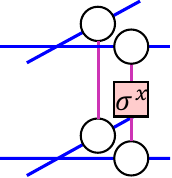}}
&=&\quad
\adjustbox{valign=m}
{\includegraphics[width=0.1\linewidth]{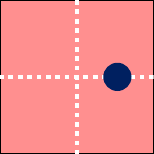}}
\text{ },
\\
&\adjustbox{valign=m}
{\includegraphics[width=0.2\linewidth]{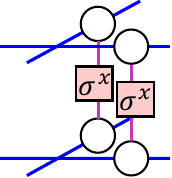}}
&=&\quad
\adjustbox{valign=m}
{\includegraphics[width=0.1\linewidth]{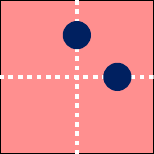}}
\text{ }.
\end{aligned}
\end{equation} The expectation value of an $R_{1}\times R_{2}$ Wilson loop by the
multiplication of matrices constructed of rows of traced PEPO tensors:

\begin{equation}\label{eq:wilson_transfer}
\begin{aligned}
\bk{\hat{W}} 
 = 
\adjustbox{valign=m}{\includegraphics[width=0.7\linewidth]{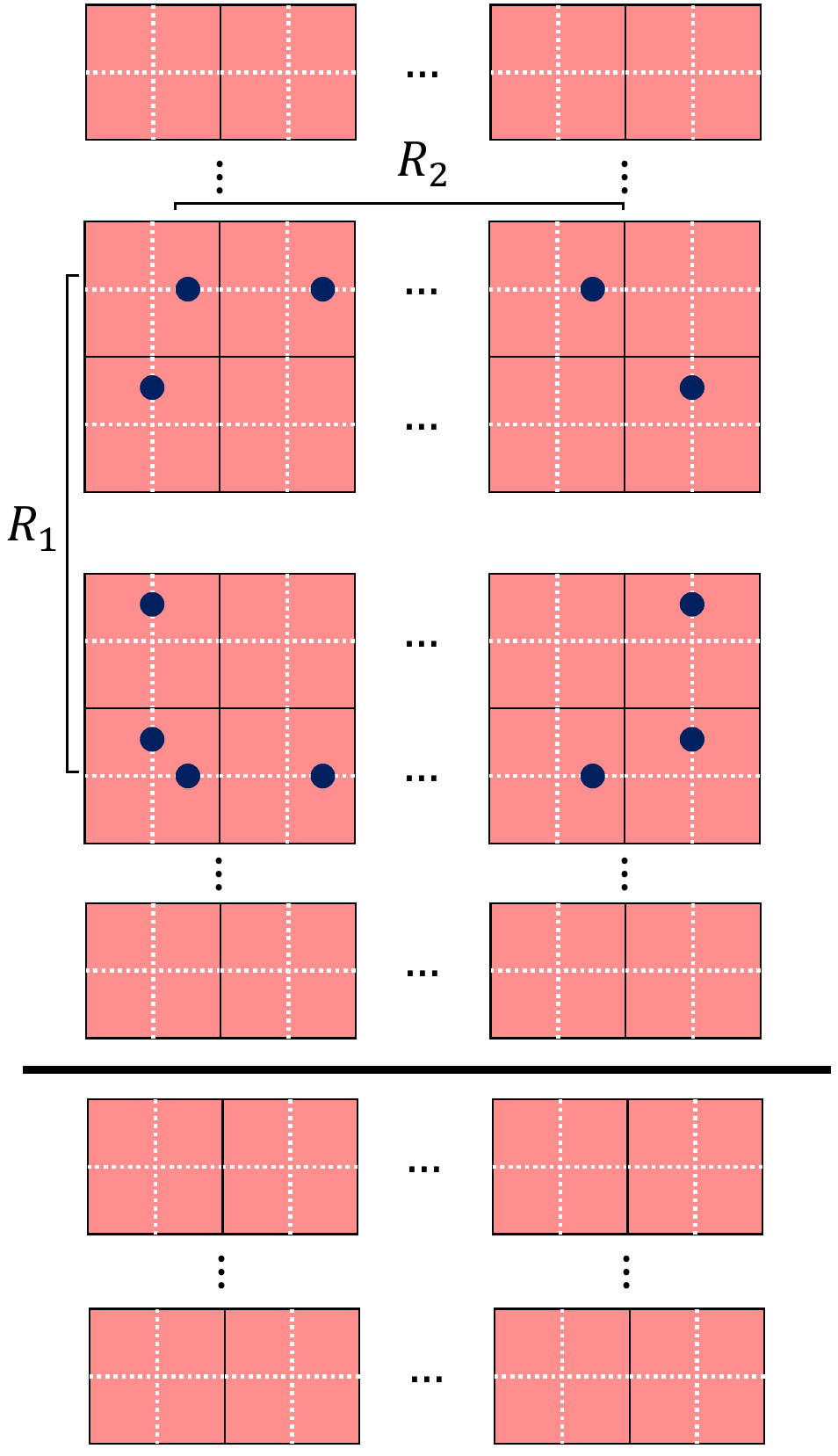}}
.
\end{aligned}
\end{equation} Based on Eq. (\ref{eq:wilson_transfer}), we define the transfer
matrices composed of contracted rows of PEPO tensors:

\begin{equation}\label{eq:E}
\begin{aligned}
&E 
& = &
\adjustbox{valign=m}{\includegraphics[width=0.7\linewidth]{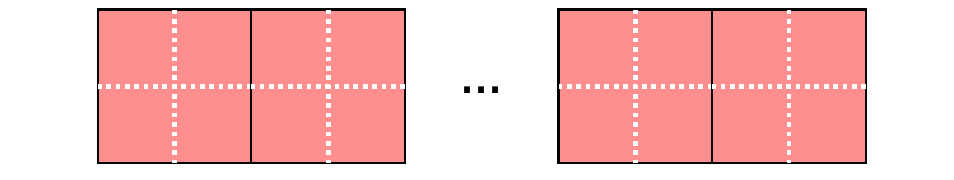}}
,
\\
&E_{||} (R)
& = &
\adjustbox{valign=m}{\includegraphics[width=0.7\linewidth]{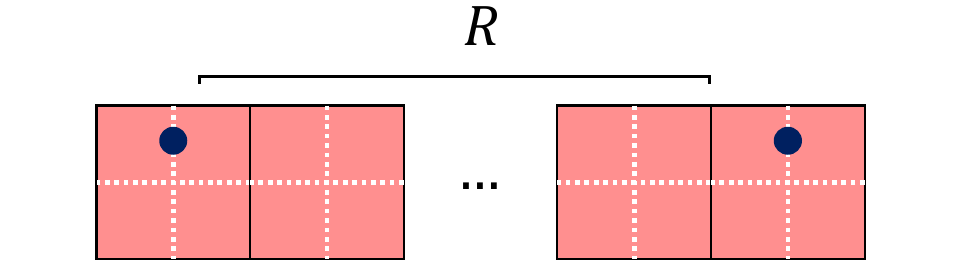}}
.
\end{aligned}
\end{equation}We briefly analyze the dependence of $\bk{\hat{W}}$ on $R_{1},R_{2}$
as is done in Ref. \citep{zoharWilsonLoopsArea2021}. The largest
eigenvalues of $E,E_{||}(R_{2})$ are denoted by $r_{1},r_{1}^{\prime}(R_{2})$
, respectively. For $R_{1}\gg1$, a perimeter-law contribution to
$\bk{\hat{W}}$ stems from a factor of $\left(r_{1}^{\prime}(R_{2})/r_{1}\right)^{R_{1}}$
due to the repetition of $E_{||}$ and $E$ in the numerator and denominator,
respectively. By requiring consistency of the computation when rotating
the system by $\pi/2$, Ref. \citep{zoharWilsonLoopsArea2021} obtains
an identical dependence of $\bk{\hat{W}}$ on $R_{1}.$ 

Ref. \citep{knauteEntanglementConfinementLattice2024} shows that
the dependence of $r_{1}^{\prime}(R)$ must be exponential:
\begin{equation}
r_{1}^{\prime}(R)=\Gamma e^{-\kappa R},\label{eq:area-law-kappa}
\end{equation}
where $\Gamma,\kappa$ are constants. $\kappa>0$ would result in
an area-law contribution to $\bk{\hat{W}}$ \textemdash{} confinement.
Therefore, we define $\kappa$ to be the measure of confinement in
our numerical computations below. $r_{1}^{\prime}(R)$ is computed
using the TEBD method for non-unitary evolutions \citep{TEBD03,TEBD07,TEBD08}
for several values of $R$, and $\kappa$ is obtained from fitting
the data as a function of $R$.

Ref. \citep{knauteEntanglementConfinementLattice2024} brings up a
similar mechanism for the study of R\'enyi entropies, and specifically,
the second RDM moment 
\begin{equation}
p_{2}=\Tr\left(\rho_{A}^{2}\right)=\exp\left(-S_{A}^{(2)}\right),\label{eq:purity}
\end{equation}
also known as the purity. The computation of the purity of an $R_{1}\times R_{2}$
rectangular system may be done by using two copies of the PEPS density
matrix. A single row of the TN construction, analogously to the Wilson
loop computation, would be:

\begin{equation}\label{eq:E_purity}
\begin{aligned}
&\tilde{E} 
 & = &\quad
\adjustbox{valign=m}{\includegraphics[width=0.35\linewidth]{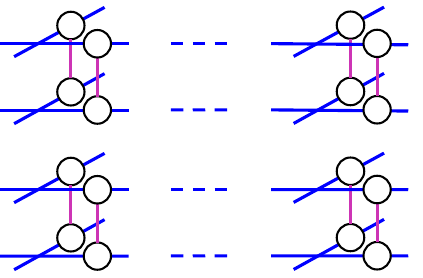}}
,
\\
&\tilde{E}_{||} (R)
 & = &
\adjustbox{valign=m}{\includegraphics[width=0.7\linewidth]{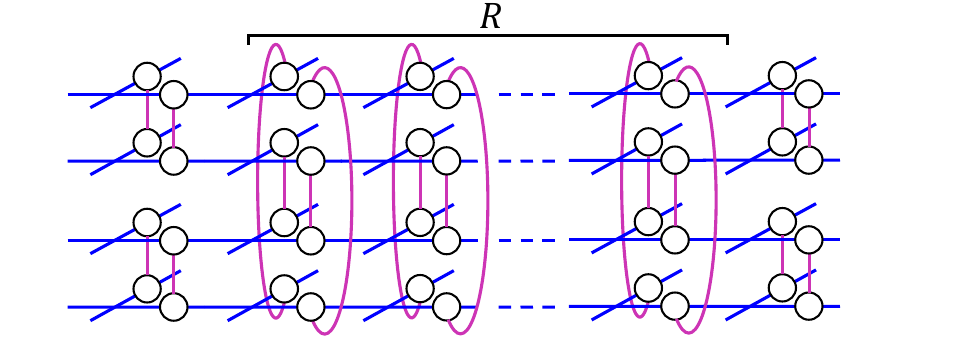}}
\\
&& = &
\adjustbox{valign=m}{\includegraphics[width=0.7\linewidth]{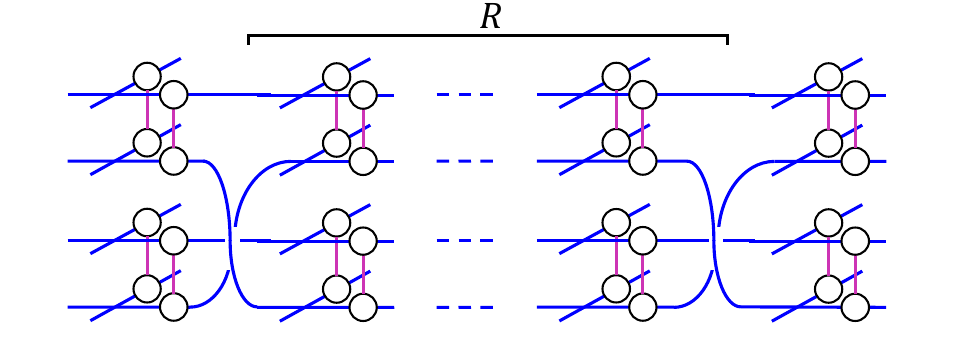}}
.
\end{aligned}
\end{equation}One may compute the SR-resolved purity of the block containing no
flux on all star-parts. If the reasonable assumption of entanglement
equipartition applies, $p_{2}\left(\left\{ \uparrow,\uparrow\dots\right\} \right)$
is equal to all other SR purities. The SR purities are obtained by
applying the local projection on the bond indices:

\begin{equation}\label{eq:E_purity_proj}
\begin{aligned}
&\tilde{E}^{\phi}\left(R\right)
  = \quad
\adjustbox{valign=m}{\includegraphics[width=0.35\linewidth]{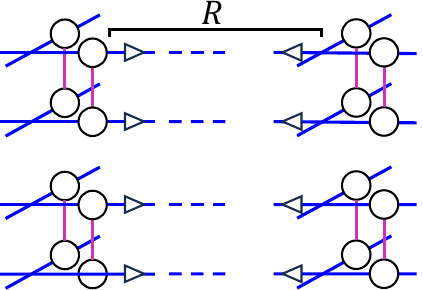}}
,
\\
&\tilde{E}^{\phi}_{||} \left(R\right)
  = 
\adjustbox{valign=m}{\includegraphics[width=0.66\linewidth]{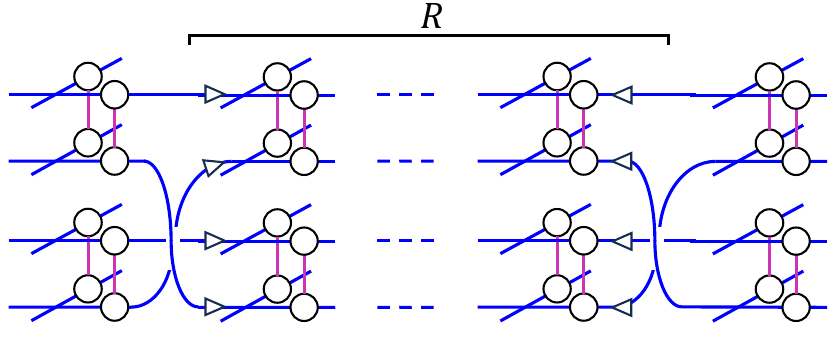}}.
\end{aligned}
\end{equation}Following Ref. \citep{zoharWilsonLoopsArea2021}, we perform a similar
analysis as in the Wilson loop case above. The largest eigenvalues
of $\tilde{E},\tilde{E}_{||}(R_{2})$ are denoted by $\tilde{r}_{1},\tilde{r}_{1}^{\prime}(R_{2})$.
In the entanglement case, the PEPS construction imposes that $p_{2},p_{2}\left(\left\{ \uparrow,\uparrow,\dots\right\} \right)$
has only area (boundary)-law contributions, therefore it is deduced
that $\tilde{r}_{1}^{\prime}(R)=\text{const.}$ always. We quantify
the area-law by $\left(\tilde{r}_{1}^{\prime}/\tilde{r}_{1}\right)$,
and expect that in the case of a corner-law entanglement, $\left(\tilde{r}_{1}^{\prime}/\tilde{r}_{1}\right)=1$.
We denote 
\begin{equation}
\eta_{d}(R)=-\log_{d}\left(\tilde{r}_{1}^{\prime}(R)/\tilde{r}_{1}\right)\label{eq:entanglement-area-law-measure}
\end{equation}
 as the area-law measure.

\subsection{Numerical results\label{subsec:Numerical-results}}

\begin{figure}[!tph]
(a)

\includegraphics[width=0.9\linewidth]{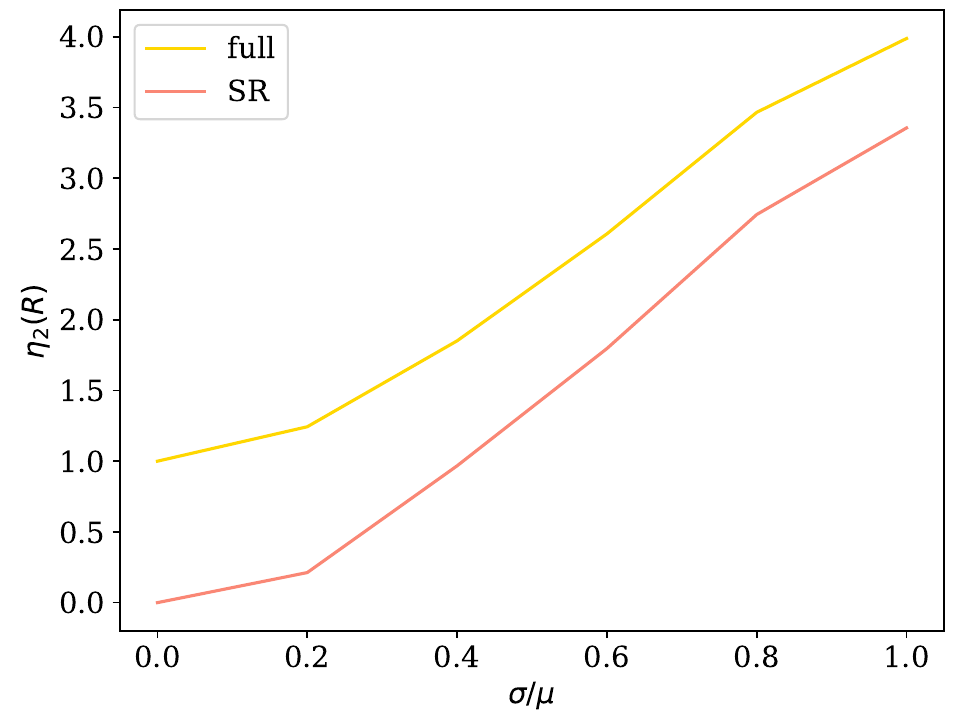}

(b)

\includegraphics[width=0.95\linewidth]{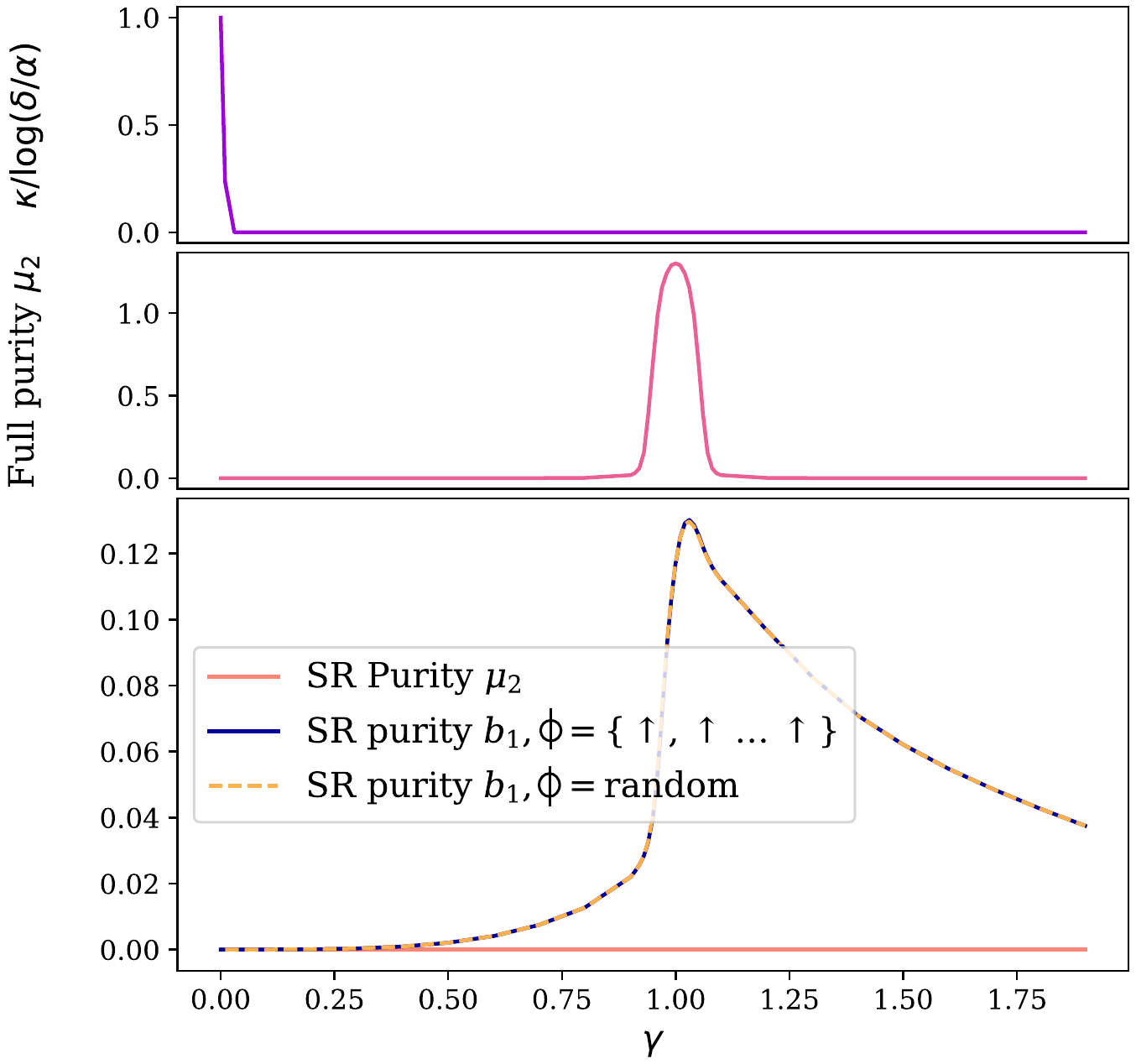}

\caption{\label{fig:results}Numerical study of full and SR entanglement in
pure $\mathbb{Z}_{2}$ gauge models. (a) Random tensors with $D=2d=4$.
The SR entanglement's dependence on the area (boundary) vanishes when
$\sigma=0$ and the state may be represented with $D=d$, and grows
as the tensor becomes far from it. (b) The generic $D=d=2$ model,
Eq. (\ref{eq:zohar_model}). The confinement measure $\kappa$ is
presented in the top panel, the full entanglement's area-law dependence
is in the middle panel. and in the bottom panel we present the area-
and corner-law dependence of the SR entanglement. Note that the behavior
of the SR entanglement is very similar, consistent with equipartition.}
\end{figure}

We start by demonstrating the corner-law entanglement by studying
a $D>d$ model. We simulate a pure $\mathbb{Z}_{2}$ gauge model with
$D=4$. The iPEPS nodes obey the restriction in Eq. (\ref{eq:allowed_terms}),
and the nonzero elements are chosen independently and randomly from
a Gaussian distribution $N(\mu,\sigma)$. We fix the mean $\mu=1$.
When $\sigma=0$, Kitaev's toric code ground state is obtained:
\[
\k{\psi}=\prod_{p}\frac{1+X^{p}}{\sqrt{2}}\k{\uparrow}^{\otimes N},
\]
where $p$ stands for plaquettes of the lattice as demonstrated in
Fig. \ref{fig:lattice}, $X^{p}=\otimes_{e\in p}\sigma_{e}^{x}$,
and $N$ is the size of the system. This model may be represented
by $D=d=2$ by choosing $\alpha=\beta=\gamma=\delta=1$ in the model
defined in Eq. (\ref{eq:zohar_model}), and is therefore expected
to have no area-law contribution to the SR entanglement; only corner-law
contribution is allowed \footnote{In fact, the corner-law contribution also vanishes in the studied
case, due to the completely local nature of the state's properties.}. As $\sigma$ is increased, the structure is expected to be lost,
and an area-law SR entanglement is observed. We compute the dependence
of the SR and full purities on system size as explained in Sec. \ref{subsec:Transfer-matrix-method}.
In Fig. \ref{fig:results}a, $\eta_{2}(R)$ is presented for th\textcolor{black}{e
full and SR purity for an open system of width $2R$ where $R=14$,
as computed by averaging over 10 occurrences of the random iPEPS tensors.
Indeed, the area-law measure $\eta_{2}$ of Eq. (\ref{eq:entanglement-area-law-measure})
vanishes in the SR purity when $\sigma=0$, and becomes larger as
$\sigma$ increases. As a result, the full entanglement's dependence
on area also grows, where when $\sigma=0$, the expected value $\eta_{2}=1$
is obtained.}

\textcolor{black}{}

\textcolor{black}{We then move to study a model with $d=D$, as defined
in Eq. (\ref{eq:zohar_model}). We choose the parameters $\beta\ll\delta\lesssim\alpha$,
in which, following Ref. \citep{zoharWilsonLoopsArea2021}, we expect
to see a confined phase at $\gamma=0$ and deconfined phase at $\gamma>0$.
Specifically, we choose $\alpha=1,\beta=0.3,\delta=0.9$ and $0\le\gamma\le2$.
First, the confinement of the phase is computed as explained in Sec.
\ref{subsec:Transfer-matrix-method}. $r_{1}^{\prime}(R)$ of Eq.
(\ref{eq:area-law-kappa}) is computed for $E_{||}(R)$ (Eq. (\ref{eq:E})),
as }defined in an open system of 64 sites, for $R=14\dots32$. $\kappa$
is then extracted and plotted in Fig. \ref{fig:results}b. As expected,
when $\gamma\ll1$, $\kappa>0$ and the phase is confined, and $\kappa=0$
deconfinement is observed for all other values of $\gamma$. As computed
in Ref. \citep{zoharWilsonLoopsArea2021}, we obtain $\kappa(\gamma=0)=-2\log\left[\delta/\alpha\right]$.

As a sanity check, we compute the dependence of $p_{2,}p_{2}\left(\left\{ \uparrow,\uparrow\dots\right\} \right)$
of rectangular systems for $R=6\dots14,$ and compute the area-law
SR entanglement as discussed in Sec. \ref{subsec:Transfer-matrix-method}.
The results are plotted in Fig. \ref{fig:results}b. As expected,
there is no dependence of the SR entanglement on the area in any value
of $\gamma$. 

\begin{figure}[!tph]
\includegraphics[width=0.65\linewidth]{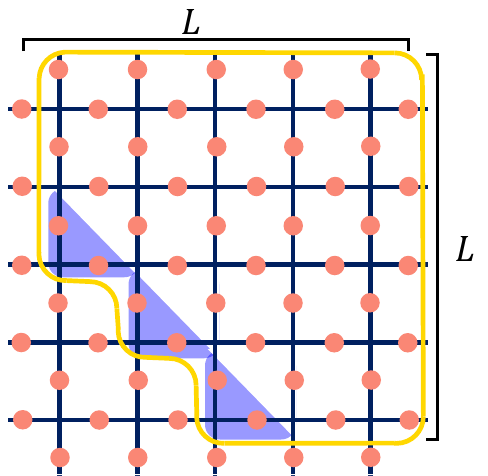}

\caption{\label{fig:stairs}The geometry we used for studying the corner-dependence
of a $D=d$ model. The number of corners (three in the illustration)
may be changed while keeping the area (boundary) of the system constant.
\textcolor{black}{Note that the bipartition was chosen such that all
corners but the bottom-left ones divide the stars into odd-sized star
parts, and therefore they do not contribute to the entanglement.}}
\end{figure}

We then study the SR entanglement as a function of the number of corners
in the system. We study a stairs-like system geometry which allows
for a constant area (boundary) length for a different number of corners,
as illustrated in Fig. \ref{fig:stairs}. \textcolor{black}{Note that
the bipartition is chosen such that not all corners contribute to
the entanglement, as can beseen in Fig. \ref{fig:stairs}.}\textcolor{blue}{{}
}In our computation, the height and width of the system is $L=6$,
and the number of corners is $c=1\dots6$ accordingly. The computation
was performed with the BMPS method \citep{BMPS}, simulating an infinite
system with open boundary conditions. The SR purity is computed for
two blocks \textendash{} The block corresponding to $s_{A}=1$ in
all star-parts, $\phi=\left\{ \uparrow,\uparrow\dots\right\} $, and
one block with a random flux distribution on the star parts. Since
$d=D$ in this model, we expect that 
\begin{align}
p_{2}\left(\phi\right) & \sim\exp\left(-\left(b_{1}c+b_{0}\right)\right),\label{eq:corner-law}
\end{align}
where $b_{0},b_{1}$ are constants. In Fig. \ref{fig:results}b, we
plot the corner-law measure $b_{1}$ for both blocks. Both blocks
display a virtually-identical behavior, supporting the applicability
of equipartition in this model. It is interesting to note that the
behavior of the entanglement, displaying a large entanglement around
$\gamma=1$ and small entanglement (SR and full) far from it. The
dependence of the SR entanglement on $\gamma$ is in no correlation
with confinement in this model. As mentioned above, the relation of
SR entanglement to confinement (or lack thereof) requires further
study.

\section{Conclusion and future outlook}

We have used the tensor network formalism to study superselection-resoved
(symmetry-resolved) entanglement in gauge invariant models. SR entanglement
is particularly interesting in such models, since it is the only accessible
(distillable) entanglement when the gauge-invariance is strictly imposed
in a system's partition. We have shown that when the state is representable
by PEPS with $D=d$, which is the case in a variation of interesting
states, the SR entanglement is bounded by a corner-law, which implies
a constant entanglement in most interesting geometries (such as rectangular
or cylindrical subsystems).

Our argument may be readily used for different lattice types, nonabelian
gauge symmetries, higher dimensions (in which an edge-law is obtained),
and models with matter. However, the variety of effects and phases
in the cases mentioned above has not been thoroughly studied. It would
be interesting to explore the range of phases and models representable
by $D=d$ and the implications of our analysis on such systems. 

Lastly, as we see in several past studies as well as our own, it seems
that the relation between confinement and entanglement, SR or general,
requires further examination. We hope that our work contributes to
future studies of this relation.
\begin{acknowledgments}
Our work has been supported by the Israel Science Foundation (ISF)
and the Directorate for Defense Research and Development (DDR\&D)
grant No. 3427/21 and by the US-Israel Binational Science Foundation
(BSF) Grant No. 2020072. NF is supported by the Azrieli Foundation
Fellows program. EZ acknowledges the support of the Israel Science
Foundation (grant No. 523/20). JK is supported by the Israel Academy
of Sciences and Humanities \& Council for Higher Education Excellence
Fellowship Program for International Postdoctoral Researchers. 
\end{acknowledgments}

\appendix

\section{Superselection-resolved entanglement in perfectly confined and deconfined
states\label{sec:appendix-Superselection-resolved-entangle-1}}

In this section, we discuss the relation between the SR entanglement
and confinement by examining the SR entanglement in states that display
an exact area- or perimeter-law decay of Wilson loop expectation value.
We show that in both cases, the SR entanglement converges to a constant
value in the limit of large subsystems. From this, it is apparent
that non-trivial SR entanglement comes from the behavior of the Wilson
loop expectation values in finite loop size, or from effects unrelated
to flux and confinement. It would be interesting to fully characterize
the relation of SR entanglement and confinement. 

\begin{figure}[h]
\includegraphics[width=0.95\linewidth]{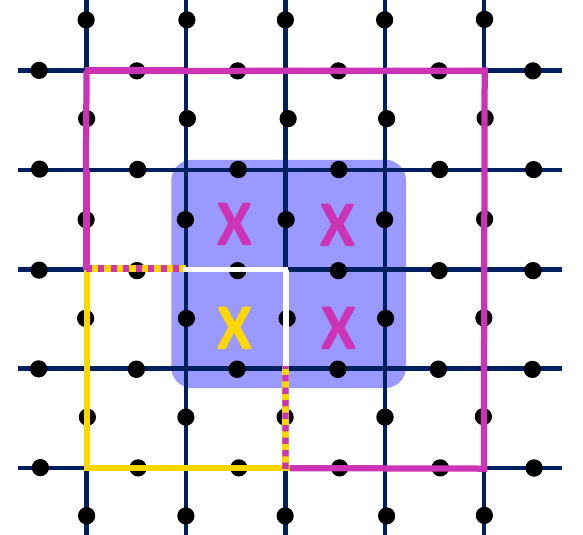}

\caption{\label{fig:directionality}The studied subsystem has two ingoing flux
lines. The directionality of the loop in the environment determines
which plaquettes in the system will contribute to the amplitude: In
yellow, the loop is closed from below the system and a single plaquette
contributes, and in pink, the loop is closed from above and three
plaquettes contribute.}
\end{figure}

We start with a perfectly-confined phase, i.e., states in which the
Wilson loops expectation value is exactly exponentially-dependent
in the Wilson loop area: 
\[
\bk{\hat{W}}=\Gamma_{a}\kappa_{a}^{\text{area}(W)},
\]
 where $\Gamma_{a},\kappa_{a}$ are constants. We observe a perfectly-confined
state of a $\mathbb{Z}_{2}$ pure-gauge field system: 
\begin{equation}
\k{\psi}=\prod_{p}\frac{1+\kappa_{a}X^{p}}{\sqrt{1+\kappa_{a}^{2}}}\k 0^{\otimes N},\label{eq:confined_state}
\end{equation}
where $p$ counts the plaquettes in the system as demonstrated in
Fig. \ref{fig:lattice} in the main text, and $X^{p}=\otimes_{e\in p}\sigma_{e}^{x}$.
For simplicity, we study the SR entanglement in blocks with a single
flux line going in and out of them, as in Fig. \ref{fig:directionality},
but the generalization to all blocks is straightforward. The state
in Eq. (\ref{eq:confined_state}), projected into the block denoted
by $\left\{ q\right\} $, may be written in the computational basis
as 
\[
\k{\psi\left(\phi\right)}=\sum_{ij}\psi_{ij}\k i_{B}\k j_{A}\propto\sum_{ij}\kappa_{a}^{\mathcal{A}(i,j)}\k i_{B}\k j_{A},
\]
 where $\mathcal{A}(i,j)$ stands for the total area of the flux loops
in the total system in the state $\k i_{B}\k j_{A}$. From Fig. \ref{fig:directionality},
we conclude that the area of the loop is determined based on the directionality
of the closing of the loop: If the loop is closed below the subsystem
(yellow route in Fig. \ref{fig:directionality}), the loop area would
be the sum of the area covered by the loop in the environment and
the area below the flux line in subsystem $A$. If the loop is closed
above the subsystem (pink route in Fig. \ref{fig:directionality}),
the total area would be the sum of the loop area in the environment
and the area above the flux line in $A$.

The projected state of the system may thus be written as 
\begin{align*}
\k{\psi\left(\phi\right)} & \propto\sum_{i\in\text{bottom }}\kappa_{a}^{\mathcal{A}(i)}\k i_{B}\sum_{j}\kappa_{a}^{\mathcal{A}(j)}\k j_{A}\\
 & \quad+\sum_{i\in\text{top}}\kappa_{a}^{\mathcal{A}(i)}\k i_{B}\sum_{j}\kappa_{a}^{\mathcal{A}(A)-\mathcal{A}(j)}\k j_{A},
\end{align*}
where $\mathcal{A}(A)$ is the area of subsystem $A$, that is, the
number of plaquettes in $A$. Note that adding inner loops in $A$
or in $B$ is consistent with the above description. 

We now denote 
\begin{align}
\k 0_{A} & =\frac{\sum_{j}\kappa_{a}^{\mathcal{A}(j)}\k j_{A}}{\sqrt{\sum_{j}\kappa_{a}^{2\mathcal{A}(j)}}},\nonumber \\
\k v_{A} & =\frac{\sum_{j}\kappa_{a}^{\mathcal{A}(A)-\mathcal{A}(j)}\k j_{A}}{\sqrt{\sum_{j}\kappa_{a}^{2\mathcal{A}(j)}}},\nonumber \\
\bk{0|v}_{A} & =\frac{\sum_{j}\kappa_{a}^{\mathcal{A}(A)}}{\sum_{j}\kappa_{a}^{2\mathcal{A}(j)}}=\frac{2^{\mathcal{A}(A)}\kappa_{a}^{\mathcal{A}(A)}}{\sum_{k}\binom{\mathcal{A}(A)}{k}\kappa_{a}^{2k}}=\left(\frac{2\kappa_{a}}{1+\kappa_{a}^{2}}\right)^{\mathcal{A}(A)}\equiv v_{0},\label{eq:confined_basis}\\
\k 1_{A} & =\frac{\k v_{A}-\k 0_{A}v_{0}}{\left|\k v_{A}-v_{0}\k 0_{A}\right|}.\nonumber 
\end{align}
The obtained RDM is enough to show that the SR entanglement is independent
on subsystem $A$\textquoteright s size and features: Since $\rho_{A}\left(\phi\right)$
may be written as a $2\times2$ matrix, the maximal number of nonzero
eigenvalues is 2, bounding the entanglement from above by $\left[S_{A}\left(\phi\right)\right]_{\text{max}}=\log2$.
However, for completeness, we continue the analysis of $\rho_{A}\left(\phi\right)$
below.

In the basis $\left\{ \k 0_{A},\k 1_{A}\right\} ,$ the block $\rho_{A}\left(\phi\right)$
may be written as
\begin{align*}
\rho_{A}\left(\phi\right) & \propto\left(\begin{array}{cc}
C_{b}+C_{t}v_{0}^{2} & C_{t}\sqrt{1-v_{0}^{2}}v_{0}\\
C_{t}\sqrt{1-v_{0}^{2}}v_{0} & C_{t}\left(1-v_{0}^{2}\right)
\end{array}\right),
\end{align*}
where $C_{b}=\sum\limits _{i\in\text{bottom}}\kappa_{a}^{2\mathcal{A}(i)},C_{t}=\sum\limits _{i\in\text{top}}\kappa_{a}^{2\mathcal{A}(i)}$
are the contributions of the environment to the states' amplitude
for flux lines closed above and below subsystem $A$. Assuming an
infinite system, one may assume $C_{b}=C_{t}$ (this assumption is
exact for a system with a reflection symmetry). The block $\rho_{A}\left({\color{blue}\phi}\right)$
then becomes
\[
\rho_{A}\left(\phi\right)=\frac{1}{2}\left(\begin{array}{cc}
1+v_{0}^{2} & \sqrt{1-v_{0}^{2}}v_{0}\\
\sqrt{1-v_{0}^{2}}v_{0} & \left(1-v_{0}^{2}\right)
\end{array}\right).
\]
The SR vN entropy would then be 
\begin{align*}
\overline{S_{A}}\left(\phi\right) & =-\sum_{\text{sign}=\pm}\frac{1}{2}\left(1+\text{sign}\cdot v_{0}\right)\log\left[\frac{1}{2}\left(1+\text{sign}\cdot v_{0}\right)\right]\\
 & \approx\log2+v_{0}^{2},
\end{align*}
where the approximation is applicable due to the exponential decay
of $v_{0}$ with subsystem $A$'s size, as can be seen in Eq. (\ref{eq:confined_basis}).
The entanglement therefore converges to a constant (and small) entanglement
as $A$'s size becomes larger. Note that the above analysis is identical
whether the boundary conditions of the system are open or periodic.

Secondly, we study a perfectly-deconfined phase, i.e., states with
an exact exponential-dependence of the Wilson loop expectation value
in its perimeter: 
\[
\bk{\hat{W}}=\Gamma_{p}\kappa_{p}^{\text{perimeter}(W)},
\]

where $\Gamma_{p},\kappa_{p}$ are constants. Such a state may be
obtained by adding string tension to Kitaev\textquoteright s toric
code model, as in Refs. \citep{castelnovoQuantumTopologicalPhase2008,trebstBreakdownTopologicalPhase2007,schuchTopologicalOrderProjected2013,orusTopologicalTransitionsMultipartite2014},
\[
\k{\psi}=\prod_{e}\frac{1+\kappa_{p}\sigma_{e}^{z}}{\sqrt{1+\kappa_{p}^{2}}}\prod_{p}\frac{1+X^{p}}{\sqrt{2}}\k 0^{\otimes N},
\]
where $e$ runs over all of the edges (gauge sites) in the system.
Here, one may quickly see that the SR entanglement between two subsystems
vanishes: The state of the system, projected into a specific block
$\left\{ q\right\} $, may be written as 
\begin{align*}
\k{\psi\left(\phi\right)} & \propto\sum_{ij}\left(1-\kappa_{p}\right){}^{\text{perimeter}(i)+\text{perimeter}(j)}\k i_{B}\k j_{A}\\
 & =\sum_{i}\left(1-\kappa_{p}\right)^{\text{perimeter}(i)}\k i_{B}\otimes\sum_{j}\left(1-\kappa_{p}\right)^{\text{perimeter}(j)}\k j_{A},
\end{align*}
which is a product state, disentangled by definition. We note in passing
that our results could be easily extended to a model wave function
with nontrivial values of both $\kappa_{a}$ and $\kappa_{p}$, displaying
a confined phase and small SR entanglement as in Eq. (\ref{eq:confined_state}).

\end{document}